\documentclass[12pt]{article}
\usepackage[margin=2.5cm]{geometry}
\usepackage{amsmath}
\usepackage{amssymb}
\usepackage[utf8]{inputenc}
\usepackage{bbold}
\usepackage{xcolor}
\usepackage[makeroom]{cancel}
\usepackage{mathtools}
\usepackage[colorlinks, allcolors=cyan]{hyperref}

\usepackage[backend=biber, style=phys, sorting=none, citestyle=numeric-comp, url=false, eprint=false]{biblatex}
\appto{\bibsetup}{\sloppy}

\bibliography{bibliography}

\title{Fuzzy Gravity: Four-Dimensional Gravity on a Covariant
Noncommutative Space and Unification with Internal Interactions}
\begin{document}
\author{Danai Roumelioti$^1$, Stelios Stefas$^1$, George Zoupanos$^{1,2,3}$}\date{}

\maketitle
\begin{center}
\itshape$^1$Physics Department, National Technical University, Athens, Greece\\
\itshape$^2$ Max-Planck Institut f\"ur Physik, M\"unchen, Germany\\
\itshape$^3$ Institut f\"ur Theoretische Physik der Universit\"at Heidelberg, Germany
\end{center}

\begin{center}
\emph{E-mails: \href{mailto:danai\_roumelioti@mail.ntua.gr}{danai\_roumelioti@mail.ntua.gr}, \href{mailto:dstefas@mail.ntua.gr}{dstefas@mail.ntua.gr}, \href{mailto:George.Zoupanos@cern.ch}{George.Zoupanos@cern.ch}}
\end{center}

\begin{abstract}
    In the present work we present an extended description of the covariant noncommutative space, which accommodates the Fuzzy Gravity model constructed previously. It is based on the historical lesson that the use of larger algebras containing all generators of the isometry of the continuous one helped in formulating a fuzzy covariant noncommutative space. Specifically a further enlargement of the isometry group leads us, in addition to the construction of the covariant noncommutative space, also to the suggestion of the group that should be gauged on such a space in order to construct a Fuzzy Gravity theory. As a result, we obtain two Fuzzy Gravity models, one in de Sitter and one in anti-de Sitter space, depending on the extension of the isometry group, and we discuss their spontaneous symmetry breaking leading to fuzzy versions of the noncommutative $SO(1,3)$ gravity. In addition we discuss for the first time how to introduce fermions in
the fuzzy gravity and even more importantly how to unify the constructed
noncommutative-fuzzy gravity with internal interactions based on $SO(10)$ or $SU(5)$ as grand unified theories.
\end{abstract}

\section{Introduction}
Noncommutativity has opened exciting new avenues both in mathematics and physics. It inspired mathematicians and physicists and brought them together in a fruitful interdisciplinary interaction. The idea of noncommutativity was firstly introduced in quantum mechanics. From the beginning, the quantum theory challenged the classical notions of commutativity. The Heisenberg uncertainty principle fundamentally altered our understanding of physics at the scales of quantum systems. It introduced the new, at the time, notion that certain pairs of physical observables, such as position and momentum, may not be compatible and should be described by noncommuting operators. Noncommutative geometry extended this idea by transferring it in the description of spaces. The mathematical developments in this new arena have profound implications, in particular when applied back to physical systems. Eventually it turns out that noncommutative geometry provided a powerful framework for addressing some of the most important problems addressed in modern physics. Together with more recent developments in physics, such as those known as fuzzy, twisted, generalized, higher, and graded geometries, noncommutative geometry found applications in a variety of areas, including the Standard Model of Particle Physics, Quantum Field Theory, Quantum Gravity, Superstring Theory, Cosmology, and Condensed Matter Physics \cite{noncomtomos}. Here we shall concentrate on Fuzzy Gravity (FG).

Searching on the possible use in Physics of the notion of noncommutative geometry, by considering the coordinates as noncommutative quantities, it can be linked to a potential quantum structure that might occur at very small distances (Planck scale). In principle, the behavior of the spacetime at these scales is unknown. Therefore, it can be considered as a rather natural step to examine the noncommutative version of General Theory of Relativity (GR), with the ambition that the latter would provide new insights particularly in regions where the very unnatural spacetime singularity appears in the conventional GR framework \cite{Maceda:2003xr, Chamseddine:2016uef}. This noncommutative gravitational theory would consist a generalization of GR, ideal for examining higher curvature scales, in which the localization of a point would be impossible to occur. Therefore, in the case of high-scales phenomena, the conventional notion of coordinates breaks down and should be substituted by elements of a noncommutative algebra.


It is well known that traditionally GR is formulated geometrically. However, there also exists an alternative gauge-theoretic approach to gravity \cite{utiyama, kibble1961, stellewest, macdowell, Ivanov:1981wn, Ivanov:1981wm, Kibble:1985sn, KAKU1977304, Fradkin:1985am, freedman_vanproeyen_2012, chamseddine_phd, CHAMSEDDINE197739, Witten:1988hc}, besides the geometric one, which started with Utiyama’s pioneering study \cite{utiyama}, and was subsequently evolved as a gauge theory of the de Sitter $SO(1,4)$ group, spontaneously broken by a scalar field to the Lorentz $SO(1,3)$ group \cite{stellewest}. In addition to GR, Weyl gravity has also been formulated as a gauge theory of the conformal group in four dimensions \cite{KAKU1977304, Fradkin:1985am}. In this case, part of the gauge fields spectrum is identified as the vielbein and the spin connection, which guarantees the interplay between the gauge-theoretic and geometric approaches through the first order formulation of GR. Taking into account the aforementioned gauge-theoretic formulation of gravity and integrating it to the noncommutative framework in which gauge theories are well formulated, led to the construction of models of noncommutative gravity \cite{Chamseddine:2000si, Chamseddine:2003we, Aschieri1, Aschieri2, Aschieri_2005, Ciric:2016isg, Cacciatori:2002gq, Cacciatori:2002ib, Aschieri3, Banados:2001xw}. In this direction, the Moyal–Weyl type of noncommutativity was used, the star-product approach was followed (where the noncommutative quantities are still ordinary functions but with an upgraded product) and eventually the Seiberg–Witten map was used \cite{seiberg-witten}. In addition to the latter, there also exists an alternative approach to the construction of noncommutative gravitational models, which makes use of the matrix-realization of the noncommutative quantities \cite{ishibasi-kawai, Aoki:1998vn, Hanada:2005vr, Furuta:2006kk, Yang:2006dk, Steinacker:2010rh, Kim:2011cr, Nishimura:2012xs,Nair:2001kr, Abe:2002in, Valtancoli:2003ve, Nair:2006qg,Banks:1996vh}. Finally, another interesting approach of noncommutative gravity, different from the ones mentioned above, can be found in refs \cite{buric-grammatikopoulos-madore-zoupanos, Buric:2007zx, Buric:2007hb}.

Next let us present our gauge-theoretic construction of the 4-dimensional matrix model of noncommutative gravity that has been done so far and how we plan to develop it further. In order to build a noncommutative gauge theory, a background noncommutative space is required to accommodate it (for a review see e.g. \cite{Manolakos:2022universe}). Therefore, first we focus on the construction of a four-dimensional covariant noncommutative space, which plays the role of the background space and then we present the gravity model, constructed as a noncommutative gauge theory on the above space.


It should be stressed that formulating gravity in the noncommutative framework is a quite difficult task given that the most obvious and common noncommutative deformations break Lorentz invariance. However, there exist certain types of deformations that constitute the covariant noncommutative spacetimes which preserve the spacetime isometries. The basic idea is to identify coordinates with the generators of a Lie algebra, in principle related to the isometries of the space time. In this way it is naturally expected to obtain coordinate noncommutativity and hopefully covariance. Snyder \cite{Snyder:1946qz} already in 1947 was the first to associate the position operators with elements of a Lie algebra, and constructed a Lorentz invariant but discrete spacetime due to the introduced unit length in his construction. Snyder’s position operators belong to the four-dimensional de Sitter algebra, $SO(1,4)$. Then Yang \cite{yang1947} in order to allow continuous translations in Snyder’s framework introduced $SO(1,5)$ and finally Kastrup \cite{kastrup_1966} considered the four-dimensional conformal algebra $SO(2,4)$. In the recent years the above early works were reconsidered in the framework of noncommutative geometry, with possible applications in string theory, by Heckman-Verlinde \cite{Heckman_2015}, where the authors build a general conformal field theory defined on covariant noncommutative versions of four-dimensional de Sitter ($dS$) or anti-de Sitter ($AdS$) spacetime. Therefore, covariant noncommutative spacetimes appear to overcome the problem of breaking Lorentz invariance, with which were confronted usually the construction of the most common noncommutative spaces. They preserve Lorentz invariance providing moreover a short-distance cutoff. In our constructions we go a natural step further, promoting the Lorentz invariance of the covariant noncommutative spacetime to a local symmetry and we produce the corresponding gravity theory on such a space.


A very interesting and relatively simple noncommutative space is a fuzzy sphere, $S^2_F$, which is the discrete matrix approximation of the ordinary, continuous, sphere, $S^2$, with the property of preserving its isometries \cite{Madore_1992}. Then clearly, its isometry group is the $SO(3)$, which is generated by the three angular momentum operators. The eigenfunctions of the corresponding Laplace operator are the well-known spherical harmonics which when are replaced by another, finite set of functions, one is led to a truncated algebra, which does not close under multiplication. In order to recover closure, the common product of these truncated functions can be upgraded to a noncommutative one, which is the matrix product. This is a way to introduce the fuzzy sphere as a matrix approximation of the ordinary one, with the coordinates defined as appropriate rescalings of the $SO(3)$ generators in a high representation. In this very interesting case the construction is straightforward, however that is not the case when a similar approach is applied in higher-dimensional spaces since the covariance is not automatically satisfied. More specifically, according to the arguments used in the fuzzy sphere case, attempt of the construction of the fuzzy four-sphere, $S^4_F$, on the same principle, would suggest to consider the $SO(5)$ group, which is corresponding isometry group, and in addition try to identify the coordinates with a subset of the generators. However, the subalgebra is not closing, and, therefore, covariance is not satisfied \cite{Kimura:2002nq}. The requirement to preserve covariance leads to employ a group with larger symmetry, in which it will be possible to do an appropriate identification of coordinates with generators, and result with a construction in which the coordinates will transform as vectors under the action of the rotational transformations. Extending minimally the symmetry leads to the adoption of the $SO(6)$ group \cite{Manolakos_paper1, Manolakos_paper2}.

The simplest covariant space is the Fuzzy Sphere, which however cannot be extended in other cases in a straightforward manner, as it is already discussed above. The suggestion resulting from the past experience is that in order to formulate a fuzzy covariant noncommutative space, one has to employ a wider symmetry containing all generators of the isometry of the continuous one. Then, in practice, in the case of the fuzzy dS four-dimensional spacetime, $dS_4$, that have been constructed \cite{Manolakos_paper1, Manolakos_paper2}, instead of using its algebra of generators, $SO(1,4)$ we consider a larger one in which the latter can be embedded. Such algebras are the $SO(1,5)$ or the conformal, $SO(2,4)$. The larger algebra, say $SO(1,5)$, contains all the generators of the subalgebra $SO(1,4)$ and the commutation relations of the latter are covariant as seen in the larger group. In this way we have successfully constructed a four dimensional FG theory \cite{Manolakos_paper1, Manolakos_paper2, Manolakos:2022universe, Manolakos:2023hif}. Let us also note that in the four-dimensional noncommutative constructions based on the idea of identifying coordinates with generators of a Lie algebra, i.e. those of Snyder \cite{Snyder:1946qz} and Yang \cite{yang1947, Heckman_2015, Manolakos_paper1, Manolakos_paper2} in the first case, using the $SO(1,4)$ it was achieved to obtain noncommutative coordinates, while in the second it was possible to moreover obtain noncommutativity of momenta and a Heisenberg-type relation, among others. In other words, the use of larger algebras provided us with more information than originally expected (recall that the motivation of Yang was only to allow continuous translations in Snyder’s framework, but much more was achieved). Therefore, it is natural for us to speculate that further enlarging the group of generators might lead to further insight of the four-dimensional FG. In particular the gauge theory used so far to construct the FG after defining the noncommutative background looks unrelated to it, since the motivation so far was the closure of the algebra of the gauge symmetry starting from the isometry group of the space, after taking into account both the commutation relations of the generators as well as the anti-commutation ones, which appear in the construction of a gauge theory in a noncommutative background. In the present work we examine the results of such a further enlargement of the Lie algebra with very encouraging findings concerning the choice of the gauge symmetry used in the construction of the FG and its relation to the noncommutative background.

Finally in the present work we discuss for the first time how to introduce fermions in the FG, an attempt which has to overcome certain difficulties, namely the chirality of the introduced fermions and their formulation in matrix representation. In addition in the present work we show how the unification of  FG with Internal Interactions in the form of Grand Unified Theories (GUTs) can be achieved.

\section{Brief Historical Review}
In this section, we shall briefly describe the lines of the logic
presented in the works of Snyder \cite{Snyder:1946qz} and Yang \cite{yang1947}, with the aim that the
motivation behind the current work becomes clearer.

\subsection*{\textit{Snyder - The \texorpdfstring{$SO(1,4)$}{SO(1,4)} case}}
As it was mentioned in the Introduction, Snyder was the first to consider a non-continuous (quantized) spacetime which preserves the usual special-relativistic Lorentz symmetries, by introducing a natural unit of length (minimal length) and assigning the spacetime coordinates to elements of the Lie algebra of the $dS$ group \cite{Snyder:1946qz}.

More specifically, the author begins by considering the four-dimensional $dS$ group, $SO(1,4)$, whose generators obey the following Lie algebra:
\begin{equation}
     \left[J_{\mu \nu},J_{\rho \sigma}\right]=i\left(\eta_{\mu \rho}J_{\nu \sigma}+\eta_{\nu \sigma}J_{\mu \rho}-\eta_{\nu \rho}J_{\mu \sigma}-\eta_{\mu \sigma}J_{\nu \rho}\right),
\end{equation}
where $\mu,\nu,\rho,\sigma=0,\dots,4$, $J_{\mu \nu}=-J_{\nu \mu}$, and $\eta_{\mu \nu}$ is the 5-dimensional Minkowski metric, with signature $\operatorname{diag}(-,+,+,+,+)$.
 Next, by considering the decomposition of 
 $SO(1,4)$ to its maximal subgroup, $SO(1,3)$, the above algebra yields the following three relations:
\begin{equation}
\begin{gathered}                            
    \left[J_{ij},J_{kl}\right]=i\left(\eta_{i k}J_{j l}+\eta_{j l}J_{i k}-\eta_{j k}J_{i l}-\eta_{i l}J_{j k}\right),\\
    \left[J_{i j},J_{k4}\right]=i\left(\eta_{i k}J_{j4}-\eta_{j k}J_{i4}\right),\\
    \left[J_{i4},J_{j4}\right]=i J_{ij},\\
\end{gathered}
\end{equation}
where $i,j,k,l=0,\dots,3$. Then we may convert the generators to physical quantities by setting 
\begin{equation}
\label{identifications.X,TH}
    \Theta_{ij}=\hbar J_{ij}, \, \text{and} \ X_i=\lambda J_{i4},
\end{equation}
respectively, where $\lambda$ is a natural unit of length, the following commutation relations are yielded
\begin{equation}
\begin{gathered} 
    \left[\Theta_{ij},\Theta_{kl}\right]=i\hbar\left(\eta_{i k}\Theta_{j l}+\eta_{j l}\Theta_{i k}-\eta_{j k}\Theta_{i l}-\eta_{i l}\Theta_{j k}\right),\\
    \left[\Theta_{i j},X_{k}\right]=i\hbar\left(\eta_{i k}X_j-\eta_{j k}X_i\right),\\
    \left[X_i,X_j\right]=\frac{i\lambda^2}{\hbar}\Theta_{ij}.\\
\end{gathered}
\end{equation}
from which, the noncommutativity of the coordinates becomes manifest.



\subsection*{\textit{Yang - The \texorpdfstring{$SO(1,5)$}{SO(1,5)} case}}
Based on Snyder's work, Yang examined whether would it be possible to attain continuous translations in the above framework \cite{yang1947}. Taking Snyder's framework one step further \cite{Snyder:1946qz, kastrup_1966, Heckman_2015}, the starting group was taken to be the $SO(1,5)$, whose generators obey the following Lie algebra:
\begin{equation}
     \left[J_{mn},J_{rs}\right]=i\left(\eta_{mr}J_{ns}+\eta_{ns}J_{mr}-\eta_{nr}J_{ms}-\eta_{ms}J_{nr}\right),
\end{equation}
where $m,n,r,s=0,\dots,5$, and $\eta_{m n}=\operatorname{diag}(-1,1,1,1,1,1)$. Performing the decompositions $SO(1,5)$ to its maximal subgroups up to $SO(1,3)$, i.e. $SO(1,5) \supset SO(1, 4)$ and $SO(1,4)\supset SO(1,3)$, turns the above commutation relation to the following nine:
\begin{equation}
\begin{gathered}  
    \left[J_{ij},J_{kl}\right]=i\left(\eta_{i k}J_{j l}+\eta_{j l}J_{i k}-\eta_{j k}J_{i l}-\eta_{i l}J_{j k}\right),\\
    \left[J_{i j},J_{k5}\right]=i\left(\eta_{i k}J_{j5}-\eta_{j k}J_{i5}\right),\\
    \left[J_{i5},J_{j5}\right]=i J_{ij},\\
    \left[J_{i j},J_{k4}\right]=i\left(\eta_{i k}J_{j4}-\eta_{j k}J_{i4}\right),\\
    \left[J_{i4},J_{j4}\right]=i J_{ij},\\
    \left[J_{i4},J_{j5}\right]=i \eta_{ij}J_{45},\\
    \left[J_{i j},J_{45}\right]=0,\\
    \left[J_{i 4},J_{45}\right]=-i J_{i5},\\
    \left[J_{i 5},J_{45}\right]=i J_{i4}.
\end{gathered}
\end{equation}
Next we may convert, as before, the generators to physical quantities by
setting
\begin{equation}
    \Theta_{ij}=\hbar J_{ij}, \, \text{and} \ X_i=\lambda J_{i5},
\end{equation}
and furthermore identify the momenta as
\begin{equation}
\label{identifications.P}
    P_i=\frac{\hbar}{\lambda}J_{i4},
\end{equation}
and set $h=J_{45}$. Given these identifications and the commutation relations above, we obtain:
\begin{equation}
\begin{gathered}  
    \left[\Theta_{ij},\Theta_{kl}\right]=i \hbar\left(\eta_{i k}\Theta_{j l}+\eta_{j l}\Theta_{i k}-\eta_{j k}\Theta_{i l}-\eta_{i l}\Theta_{j k}\right),\\
    \left[\Theta_{i j},P_{k}\right]=i\hbar\left(\eta_{i k}P_{j}-\eta_{j k}P_{i}\right),\\
    \left[P_{i},P_{j}\right]=\frac{i \hbar}{\lambda^2} \Theta_{ij},\\
    \left[\Theta_{i j},X_{k}\right]=i\hbar \left(\eta_{i k}X_{j}-\eta_{j k}X_{i}\right),\\
    \left[X_{i},X_{j}\right]=\frac{i\lambda^2}{\hbar} \Theta_{ij},\\
    \left[X_{i},P_{j}\right]=i\hbar \eta_{ij}h,\\
    \left[\Theta_{i j},h\right]=0,\\
   \left[X_{i},h\right]=\frac{i\lambda^2}{\hbar} P_{i},\\
    \left[P_{i},h\right]=-\frac{i\hbar}{\lambda^2} X_{i}.
\end{gathered}
\end{equation}
From the above relations, it becomes clear that extending the starting group from $SO(1,4)$ to $SO(1,5)$ allows the momenta to be seamlessly incorporated into this extended algebra. This integration leads to two significant results: firstly, since the momenta are now elements of this Lie algebra, they exhibit a noncommutative behavior, implying that the momentum space itself becomes quantized. Secondly, it becomes evident that the commutation relation
between coordinates and momenta naturally yields a Heisenberg-type uncertainty relation.

\section{The background space}
Before we move on with the gauge theory of FG, we will first have to establish the background space, on which this theory will be formulated. In the current work, we will examine both the four-dimensional $dS_4$ and $AdS_4$ spacetimes. 

To begin, we shall consider the embedding of the above spaces in a five-dimensional flat one given the following embedding equation:
\begin{equation}
\label{constrainteq}
    \eta_{\mu \nu}x^{\mu} x^{\nu}=s R^2 ,
\end{equation}
where $R$ the radius of the space, $\eta_{\mu \nu}=(-1,1,1,1,s)$ and $s=\pm 1$. The groups describing the isometries of the above spaces are for $s=+1$ the $dS$ group, $SO(1,4)$, while for $s=-1$ the $AdS$ group, $SO(2,3)$. For the moment, we will not distinguish between them.

The first thing we do is to consider whether the covariance of the space is preserved, that is if there are enough generators in the chosen group to accommodate 5 coordinates as well as $\frac{5\cdot 4}{2}=10$ generators to form a Lorentz subgroup, all together as generators of the group. Since $SO(1,4)$ (just as $SO(2,3)$) only has 10 generators, it is obvious that the covariance of the space is not preserved, as the smallest number of generators needed is 15. For reasons of covariance we first consider the minimal extension of $SO(1,4)$ to $SO(1,5)$ (or $SO(2,3)\subset SO(2,4)$ for the $AdS$ case), which nicely has 15 generators. Having in mind the logic presented in the previous section, we shall further consider the extension of $SO(1,5)$ to $SO(1,6)$ (or $SO(2,4) \subset SO(2,5)$), in order to determine whether we will obtain any additional results\footnote{In the same way the extension of $SO(1,4)$ to $SO(1,5)$ manifested the noncommutativity of momentum as well as the Heisenberg-type uncertainty relation.}. Finally, in order to reach a four-dimensional description, we shall consider the three-step decomposition $SO(1,6) \supset SO(1,5) \supset SO(1,4) \supset SO(1,3)$ (or $SO(2,5) \supset SO(2,4) \supset SO(2,3) \supset SO(1,3)$). 

The $SO(1,6)$ group (and similarly the $SO(2,5)$) comprises 21 generators $J_{MN}$ (with $M,N=0,\dots,6$), which obey the following commutation relations:
\begin{equation}
\left[J_{MN},J_{RS}\right]=i\left(\eta_{MR}J_{NS}+\eta_{NS}J_{MR}-\eta_{NR}J_{MS}-\eta_{MS}J_{NR}\right),
\end{equation}
where $\eta_{MN}=\operatorname{diag}(-1,1,1,1,s,1,1)$.

The non-zero commutation relations that arise from above algebra after the aforementioned decompositions are the following:
\begin{equation}
\begin{gathered}  
    \left[J_{ij},J_{kl}\right]=i\left(\eta_{i k}J_{j l}+\eta_{j l}J_{i k}-\eta_{j k}J_{i l}-\eta_{i l}J_{j k}\right),\\
    \left[J_{i j},J_{k6}\right]=i\left(\eta_{i k}J_{j6}-\eta_{j k}J_{i6}\right),\\
    \left[J_{i j},J_{k5}\right]=i\left(\eta_{i k}J_{j5}-\eta_{j k}J_{i5}\right),\\
    \left[J_{i j},J_{k4}\right]=i\left(\eta_{i k}J_{j4}-\eta_{j k}J_{i4}\right),\\
    \left[J_{i6},J_{j6}\right]=i J_{ij},\\
    \left[J_{i6},J_{j5}\right]=-i \eta_{ij}J_{56},\\
    \left[J_{i6},J_{j4}\right]=-i \eta_{ij}J_{46},\\
    \left[J_{i6},J_{56}\right]=iJ_{i5},\\
    \left[J_{i6},J_{46}\right]=iJ_{i4},\\
    \left[J_{i5},J_{j5}\right]=i J_{ij},\\
    \left[J_{i5},J_{j4}\right]=-i \eta_{ij}J_{45},\\
    \left[J_{i 5},J_{56}\right]=-i J_{i6},\\
    \left[J_{i 5},J_{45}\right]=i J_{i4},\\
    \left[J_{i4},J_{j4}\right]=i s J_{ij},\\
    \left[J_{i 4},J_{46}\right]=-i s J_{i6},\\
    \left[J_{i 4},J_{45}\right]=-i s J_{i5},\\
    \left[J_{56},J_{46}\right]=-i J_{45},\\
    \left[J_{56},J_{45}\right]=i J_{46},\\
    \left[J_{46},J_{45}\right]=-i s J_{56},
\end{gathered}
\end{equation}
where $\eta_{ij}=\operatorname{diag}(-1,1,1,1)$.

Having the initial algebra written in this $SO(1,3)$ language, we identify the noncommutativity tensor, the coordinates and momenta as:
\begin{equation}
    \Theta_{ij}=\hbar J_{ij}, \,  X_i=\lambda J_{i5},\  \text{and} \ P_i=\frac{\hbar}{\lambda}J_{i4}.
\end{equation}
Furthermore, we identify the remaining generators as:
\begin{equation}
    Q_i=\frac{\hbar}{\lambda}J_{i6}, \,  q= J_{56}, \,  p= J_{46}, \ \text{and} \ h=J_{45}.
\end{equation}

Admitting the above identifications, the algebra now becomes:
\begin{equation}
\label{backgroundAlgebra}
\begin{gathered}  
    \left[\Theta_{ij},\Theta_{kl}\right]=i\hbar\left(\eta_{i k}\Theta_{j l}+\eta_{j l}\Theta_{i k}-\eta_{j k}\Theta_{i l}-\eta_{i l}\Theta_{j k}\right),\\
    \left[\Theta_{i j},Q_{k}\right]=\frac{i}{\hbar} \left(\eta_{i k}Q_{j}-\eta_{j k}Q_{i}\right),\\
    \left[\Theta_{i j},X_{k}\right]=\frac{i}{\hbar}\left(\eta_{i k}X_{j}-\eta_{j k}X_{i}\right),\\
    \left[\Theta_{i j},P_{k}\right]=\frac{i}{\hbar}\left(\eta_{i k}P_{j}-\eta_{j k}P_{i}\right),\\
    \left[Q_{i},Q_{j}\right]=i\frac{\hbar}{\lambda^2} \Theta_{ij},\\
    \left[Q_{i},X_{j}\right]=-i\frac{\hbar}{\lambda^2} \eta_{ij}q,\\
    \left[Q_{i},P_{j}\right]=-i\frac{\hbar^2}{\lambda^2} \eta_{ij}p,\\
    \left[Q_{i},q\right]=i\frac{\hbar}{\lambda^2}X_{i},\\
    \left[Q_{i},p\right]=iP_{i},\\
    \left[X_{i},X_{j}\right]=i \frac{\lambda^2}{\hbar}\Theta_{ij},\\
    \left[X_{i},P_{j}\right]=-i \hbar\eta_{ij}h,\\
    \left[X_{i},q\right]=-i \frac{\lambda^2}{\hbar}Q_{i},\\
    \left[X_{i},h\right]=i\frac{\lambda^2}{\hbar} P_{i},\\
    \left[P_{i},P_{j}\right]=i s \frac{\hbar}{\lambda^2}\Theta_{ij},\\
    \left[P_{i}, p\right]=-i s Q_{i},\\
    \left[P_{i}, h\right]=-i s \frac{\hbar}{\lambda^2}X_{i},\\
    \left[q,p\right]=-i h,\\
    \left[q, h\right]=i p,\\
    \left[p, h\right]=-i s q.
\end{gathered}
\end{equation}

\section{Noncommutative gauge theory of 4D gravity}
\subsection{Gauge group and representation}
Starting with the formulation of a gauge theory for gravity in the space mentioned above, we first have to determine the group that will be gauged.
 Naturally, the group we will choose is the one that describes the symmetries of the theory, in this case, the isometry group of $dS_4$, $SO(1,4)$ (or in the $AdS_4$ case, $SO(2,3)$).

Since, as it is shown in \cite{Manolakos_paper1,Manolakos_paper2}, in noncommutative gauge theories the use of anticommutators of the algebra generators is inevitable, and since the anticommutators of the generators of either of the above isometry groups do not necessarily yield elements that belong in the algebra themselves, we have to take into account the closing of the anticommutators of the relevant gauge group. In other words, we need all the results of the anticommutators to belong to the generators of the chosen gauge group. In order to achieve that, we will have to pick a specific representation of the algebra generators, and subsequently extend the initial gauge group to one with larger symmetry, in which both the commutator and anticommutator algebras close. Following this procedure, we are led to the extension of our initial gauge groups $SO(1, 4)$ and $SO(2, 3)$ to the $SO(1, 5)\times U(1)$ in the $dS_4$ case, and to the $SO(2, 4) \times U(1)$ correspondingly in the $dS_4$. It is worth noting that the algebra \eqref{backgroundAlgebra} contains all the information about the groups $SO(1, 5)$ and $SO(2, 4)$ (with $s = +1$ and $-1$ correspondingly) that should be gauged. This is the bonus given by the extension of the isometries of the background spaces from $SO(1, 5)$ to $SO(1, 6)$ and from $SO(2, 4)$ to $SO(2,5)$. The $U(1)$ due to the anticommutators is not given by this extension.

Given the similarities of the commutation relations of the generators of the two algebras in \eqref{backgroundAlgebra} (the difference being only the sign of $s$) in the following we discuss only the gauging of the latter, $SO(2,4) \times U(1)$.

To start with, we choose to fix the representation of the generators to the standard four-dimensional Dirac representation. In this case, the generators of $SO(1,4)$ (and $SO(2,3)$) can be formed using combinations of the $\gamma$-matrices, which satisfy:
\begin{equation}
    \{\gamma_a,\gamma_b\}=-2\eta_{ab}\mathbb{1}_4,
\end{equation}
where $a,b=1,\dots,4$, $\eta_{ab}$ the (mostly positive) Minkowski metric, and $\mathbb{1}_4$ is the four-dimensional identity matrix. The representation of these matrices is chosen to be the following:
\begin{equation}\label{representation}
\begin{aligned}
\gamma_1 & =\left(\begin{array}{cccc}
0 & 0 & 1 & 0 \\
0 & 0 & 0 & 1 \\
1 & 0 & 0 & 0 \\
0 & 1 & 0 & 0
\end{array}\right), & \gamma_2=\left(\begin{array}{cccc}
0 & 0 & 0 & 1 \\
0 & 0 & 1 & 0 \\
0 & -1 & 0 & 0 \\
-1 & 0 & 0 & 0
\end{array}\right), \\ \\
\gamma_3 & =\left(\begin{array}{cccc}
0 & 0 & 0 & -i \\
0 & 0 & i & 0 \\
0 & i & 0 & 0 \\
-i & 0 & 0 & 0
\end{array}\right), & \gamma_4=\left(\begin{array}{cccc}
0 & 0 & 1 & 0 \\
0 & 0 & 0 & -1 \\
-1 & 0 & 0 & 0 \\
0 & 1 & 0 & 0
\end{array}\right), 
\end{aligned}
\end{equation}
and the $\gamma_5$ matrix is defined as $\gamma_5=\gamma_2\gamma_3\gamma_4\gamma_1$, 
\begin{equation}
\gamma_5 =\left(\begin{array}{cccc}
1 & 0 & 0 & 0 \\
0 & 1 & 0 & 0 \\
0 & 0 & -1 & 0 \\
0 & 0 & 0 & -1
\end{array}\right).
\end{equation}
Next, concentrating only in the $SO(2,4) \times U(1)$ case we identify its
generators in the representation \eqref{representation} as follows:
\begin{itemize}
    \item Six Lorentz generators $M_{ab}=-\frac{i}{4}\left[\gamma_a,\gamma_b\right]$
    \item Four generators for translations $P_a=-\frac{1}{2}\gamma_a(1-\gamma_5)$
    \item Four generators for conformal boosts $K_a=\frac{1}{2}\gamma_a(1+\gamma_5)$
    \item One generator for dilatations $D=-\frac{1}{2}\gamma_5$ and
    \item One $U(1)$ generator $\mathbb{1}_4$.
\end{itemize}

The above definitions of the generators lead to the following commutation relations (which is the well-known conformal algebra):
\begin{equation}
\label{Lcom}
\begin{aligned}
{\left[M_{a b}, M_{c d}\right] } & =\eta_{b c} M_{a d}+\eta_{a d} M_{b c}-\eta_{a c} M_{b d}-\eta_{b d} M_{a c}, \\
{\left[M_{a b}, P_c\right] } & =\eta_{b c} P_a-\eta_{a c} P_b, \\
{\left[M_{a b}, K_c\right] } & =\eta_{b c} K_a-\eta_{a c} K_b, \\
{\left[P_a, D\right] } & =P_a, \\
{\left[K_a, D\right] } & =-K_a, \\
{\left[K_a, P_b\right] } & =-2\left(\eta_{a b} D+M_{a b}\right),
\end{aligned}
\end{equation}
and the following anticommutation relations:
\begin{equation}
\label{Lanticom}
\begin{aligned}
 \{M_{a b}, M_{c d}\}&=\frac{1}{2}\left(\eta_{a c}\eta_{b d}-\eta_{b c} \eta_{a d}\right)-i\epsilon_{abcd}D, \\
\{M_{a b}, P_c\} &=+i\epsilon_{abcd}P^d, \\
\{M_{a b}, K_c\}&=-i\epsilon_{abcd}K^d, \\
\{M_{ab},D\}&= 2M_{ab}D,\\
\{P_{a}, K_{b}\}&=4M_{ab}D+\eta_{ab},\\
\{K_{a}, K_{b}\}&=\{P_{a}, P_{b}\}=-\eta_{a b},\\
\{P_{a}, D\}&=\{K_{a}, D\}=0.
\end{aligned}
\end{equation}

\section{Fuzzy gravity}
To this point, we have concluded that the four-dimensional noncommutative gravity on $AdS_4$ can be formulated as a gauge theory of the $SO(2,4)\times U(1)$ group \footnote{The noncommutativity gravity on $dS_4$ can be discussed similarly}.
To make the discussion here self-contained, we proceed as in ref \cite{Manolakos_paper1} in
order to determine the action of the theory. We may start with the
following topological action
\begin{equation}
\label{topolaction}    \mathcal{S}=\operatorname{Tr}\left(\left[X_{\mu}, X_{\nu}\right]-\kappa^{2} \Theta_{\mu \nu}\right)\left(\left[X_{\rho}, X_{\sigma}\right]-\kappa^{2} \Theta_{\rho \sigma}\right) \epsilon^{\mu \nu \rho \sigma}\,,
\end{equation}
with the following field equations:
\begin{equation}
    \epsilon^{\mu \nu \rho \sigma}\left[X_{\nu},\left[X_{\rho}, X_{\sigma}\right]-\kappa^{2} \Theta_{\rho \sigma}\right]=0\,,\,\, \epsilon^{\mu \nu \rho \sigma}\left(\left[X_{\rho}, X_{\sigma}\right]-\kappa^{2} \Theta_{\rho \sigma}\right)=0\,,
\end{equation}
which in the case where $\kappa^2=\frac{i\lambda^2}{\hbar}$, the second equation yields the equation which defines the noncommutativity of the space, and the first one is trivially satisfied.

Then the topological action can be promoted to a dynamical one, by the inclusion of the gauge fields of the theory in the action as fluctuations of the independent fields $X$ and $\Theta$ as 
\begin{equation}
\begin{gathered}
    \mathcal{S}=\operatorname{Trtr} \epsilon^{\mu \nu \rho \sigma}\left(\left[X_{\mu}+A_{\mu}, X_{\nu}+A_{\nu}\right]-\kappa^{2}\left(\Theta_{\mu \nu}+\mathcal{B}_{\mu \nu}\right)\right)\\
   \qquad \qquad \qquad \qquad \qquad \cdot \left(\left[X_{\rho}+A_{\rho}, X_{\sigma}+A_{\sigma}\right]-\kappa^{2}\left(\Theta_{\rho \sigma}+\mathcal{B}_{\rho \sigma}\right)\right)\,,
\end{gathered}
\end{equation}
where a trace `$\operatorname{tr}$' over the gauge group generators is also inserted. The last action, can nicely take a more familiar form by identifying the \textit{covariant coordinate}, $\mathcal{X}_{\mu}\equiv X_{\mu}+A_{\mu}$, where $A_{\mu}$ is the gauge connection \cite{Madore:2000en},
\begin{equation*}
    A_\mu = a_\mu \otimes \mathbb{1}_{4} +  \omega_\mu{}^{ab}\otimes M_{ab}+ e_\mu{}^{a}\otimes P_a + b_\mu{}^{a}\otimes K_a + \tilde{a}_\mu\otimes D,
\end{equation*}
the \textit{covariant noncommutativity tensor}, $\hat{\Theta}_{{\mu}{\nu}}\equiv\Theta_{{\mu}{\nu}}+\mathcal{B}_{{\mu}{\nu}}$, in which $\mathcal{B}_{{\mu}{\nu}}$ is a 2-form field taking care of the transformation of $\Theta$, and the \textit{field strength tensor} of the gauge theory $\hat{F}_{\mu \nu}\equiv\left[\mathcal{X}_{\mu}, \mathcal{X}_{\nu}\right]-\kappa^2 \hat{\Theta}_{\mu \nu}$ \cite{Madore_1992}. Since the latter is an element of the gauge algebra, it can be expanded on the algebra's generators as 
\begin{equation}
    \hat{F}_{\mu \nu}= R_{\mu \nu} \otimes \mathbb{1}_4 +\frac{1}{2} R_{\mu \nu}{}^{a b} \otimes M_{a b} + \tilde{R}_{\mu \nu}{}^{a} \otimes P_{a}+R_{\mu \nu}{}^{a} \otimes K_{a}+\tilde{R}_{\mu \nu} \otimes D\,.
\end{equation}
Consequently, in terms of the above definitions, the dynamical action can be written in the following form:
\begin{equation}\label{actionssb}
\mathcal{S}=\operatorname{Tr}\operatorname{tr} \epsilon^{\mu \nu \rho \sigma}\left(\left[\mathcal{X}_\mu,\mathcal{X}_\nu \right]-\frac{i\lambda^2}{\hbar}\hat{\Theta}_{\mu\nu}\right) \left(\left[\mathcal{X}_\rho,\mathcal{X}_\sigma \right]-\frac{i\lambda^2}{\hbar}\hat{\Theta}_{\rho\sigma}\right)\coloneq\operatorname{Tr}\operatorname{tr} \epsilon^{\mu \nu \rho \sigma}\hat{F}_{\mu \nu}\hat{F}_{\rho \sigma}\, ,
\end{equation}
which, again, after variations w.r.t. $\mathcal{X}$ and $\hat{\Theta}$ lead to the field equations 
\begin{equation}
    \epsilon^{\mu \nu \rho \sigma} \hat{F}_{\rho \sigma}=0\,,\,\, \epsilon^{\mu \nu \rho \sigma}\left[\mathcal{X}_{\nu}, \hat{F}_{\rho \sigma}\right]=0\,,
\end{equation}
the vanishing of the field strength tensor and the noncommutative analogue of the Bianchi identity respectively.

In order to spontaneously break $SO(2,4)\times U(1)$ gauge symmetry down to the Lorentz, we shall introduce a scalar field, $\Phi(X)$, belonging to the adjoint rep, $15$, of $SU(4)\sim SO(2,4)$ (or 2nd rank antisymmetric of $SO(2,4)$) in the action \eqref{actionssb}, and fix it in the gauge that leads to the Lorentz group (see \cite{Manolakos_paper1, Manolakos_paper2, Roumelioti:2024lvn}). This scalar field must be charged under $U(1)$ gauge symmetry in order for it to break, and not appear in the final broken symmetry. Introducing the scalar field, the above action takes the form:
\begin{equation}
\mathcal{S}=\operatorname{Trtr} \Big[\lambda \Phi(X) \varepsilon^{\mu \nu \rho \sigma}\hat{F}_{\mu \nu} \hat{F}_{\rho \sigma} +\eta\left(\Phi(X)^2-\lambda^{-2} \mathbb{1}_N \otimes \mathbb{1}_4\right)\Big],
\end{equation}
where $\eta$ is a Lagrange multiplier, and $\lambda$ is a dimensionfull parameter. The scalar field itself is also an element of the gauge group, and hence can be expanded on its generators, as
\begin{equation}
\Phi(X)=\phi(X) \otimes \mathbb{1}_4+\phi^{a b}(X) \otimes M_{a b}+\tilde{\phi}^a(X) \otimes P_a+\phi^a(X) \otimes K_a+\tilde{\phi}(X) \otimes D.
\end{equation}
As in our previous works, \cite{Manolakos_paper1, Manolakos_paper2, Roumelioti:2024lvn}, we gauge-fix the scalar field into the dilatation direction:
\begin{equation}
\Phi(X)=\left.\tilde{\phi}(X) \otimes D\right|_{\tilde{\phi}=-2 \lambda^{-1}}=-2 \lambda^{-1} \mathbb{1}_N \otimes D.
\end{equation}

On-shell, when the above equality is on hold, and after the consideration of the anticommutator relations of the generators, and, afterwards, the traces of the various generators' products, the action reduces to
\begin{equation}
\mathcal{S}_{b r}=\operatorname{Tr}\left(\frac{\sqrt{2}}{4} \varepsilon_{a b c d} R_{m n}{}^{a b} R_{r s}{}^{c d}-4 R_{m n} \tilde{R}_{r s}\right) \varepsilon^{m n r s},
\end{equation}
while any other term, along with the Lagrange multiplier, has vanished due to the gauge fixing. This resulting action, bearing the remaining $SO(1,3)$ gauge symmetry after the spontaneous symmetry breaking, when the noncommutative limit is considered, that is when noncommutativity vanishes (and after some reparametrizations which concern the noncommutative to commutative equivalents of the several fields and tensors), reduces to the Palatini action, which is equivalent to the Einstein-Hilbert, along with a cosmological constant term (for details see \cite{Manolakos_paper2}). In other words, ordinary GR in the presence of a cosmological constant is being retrieved.

\section{Fermions in the Fuzzy Gravity and Unification with Internal
Interactions}

Attempting to unify FG with internal interactions, say along the lines of Unification of Conformal Gravity with $SO(10)$ \cite{Roumelioti:2024lvn}, the difficulties that in principle one is facing are that fermions should (i) be chiral in order to have a chance to survive in low energies and not receive masses as the Planck scale, (ii) appear in a matrix representation, since the constructed FG is a matrix model. A way out was suggested in unification schemes with extra fuzzy dimensions \cite{Chatzistavrakidis:2010xi,Chatzistavrakidis:2011toc}.

The suggested way out can be summarized as follows: Instead of using fermions in fundamental or tensor representations of $SU(N)$ gauge groups, we can use bi-fundamental representations of cross product gauge groups. The same recipe can be extended to $SO(N)$ gauge theories. This kind of constructions was suggested earlier in various contexts and with different motivations \cite{Kachru:1998ys,Ibanez:1998xn, Ma:2004mi, Leontaris:2005ax,Irges:2011de,Manolakos:2020cco,Patellis:2024dfl}.

In \cite{Roumelioti:2024lvn} was suggested that the unification of the conformal gravity with internal interactions based on the GUT $SO(10)$ could be achieved using the $SO(2,16)$ as unifying gauge group. Recall that the whole strategy was based on the observation that the dimension of the tangent space is not necessarily equal to the dimension of the corresponding curved manifold \cite{Weinberg:1984ke, Percacci:1984ai, Percacci_1991, Nesti_2008, Nesti_2010, Krasnov:2017epi, Chamseddine2010, Chamseddine2016, Manolakos:2023hif, noncomtomos, Konitopoulos:2023wst}. An additional fundamental observation \cite{Roumelioti:2024lvn} is that on the fermions used it was permitted to impose Weyl and Majorana conditions. More specifically, using Euclidean signature for simplicity (the implications of using non-compact space are explicitly discussed in \cite{Roumelioti:2024lvn}), one starts with $SO(18)$ and with the fermions in its spinor representation, $256$. Then the spontaneous symmetry breaking of $SO(18)$ leads to its maximal subgroup $SU(4)\times SO(12)$ \cite{Roumelioti:2024lvn} i.e. 
\begin{equation}\begin{aligned}
  SO(18) &\rightarrow SO(6) \times SO(12)\\
   256 &= (4, \bar{32}) + (\bar{4}, 32).
\end{aligned}\end{equation}
Given that the Majorana condition can be imposed\footnote{According to \cite{Roumelioti:2024lvn} due to the non-compactness of the used $SO(2,16) \sim SO(18)$.}, we are led after the spontaneous symmetry breaking to the gauge theory $SO(6) \times SO(12)$ with fermions in the $(4,\bar{32})$ representation. Then note that choosing to start with the $SO(6) \times SO(12)$ as the initial gauge theory with fermions in the $(4,\bar{32})$ representation, we satisfy the criteria to obtain chiral fermions in tensorial representation of a fuzzy space, mentioned above. In addition it should be stressed that imposing Weyl and Majorana conditions do not concern the global or local nature of the gauge part of the theory.

Then, according to \cite{Roumelioti:2024lvn}, the following spontaneous symmetry breakings can be achieved by using scalars in the appropriate representations,
\begin{equation}
SO(6) \rightarrow SU(2) \times SU(2),
\end{equation}
in the conformal gravity sector ,and
\begin{equation}
  SO(12) \rightarrow SO(10) \times [U(1)]_\text{global}
\end{equation}
in the internal gauge symmetry sector, with fermions in the the $16_L (-1)$. Here the gauge symmetry $U(1)$ of the gravity part, appearing due to the anticommutation relations, is identified with the $U(1)$ appearing in the branching rule of the internal group $SO(12) \supset SO(10) \times U(1)$. The other generations are introduced as usual with more chiral fermions in the $SO(2,16)$.

It is worth noting that if it was chosen to gauge the $SO(1,5) \times U(1)$ group as an alternative to $SO(2,4) \times U(1)$, then an obvious unifying group of both FG and internal interactions would be the $SO(1,17)$, instead of the $SO(2,16)$. Then, the type of spinors obtained, governed by the signature $(p-q)\operatorname{mod}(8)$ \cite{D_Auria_2001,majoranaspinors}, in this specific case would give zero signature leading to real reps instead of complex that are required in a chiral theory. Another unifying group that one could try is the $SO(1,15)$. In that case the signature is six and is permitted to impose both Weyl and Majorana conditions and obtain complex reps. However, there is no room for the $U(1)$ required to be part of the gauge group due to the antisymmetry of the generators if we insist to have $SO(10)$ as the GUT of the internal interactions as in previous cases \cite{Chamseddine2016, Nesti_2010,Konitopoulos:2023wst,Manolakos:2023hif}. In the present context it could be a viable case if we consider $SU(5)$ as GUT. However the original Georgi-Glashow model \cite{Georgi:1974sy} has been ruled out phenomenologically long time ago, therefore it cannot be considered seriously without major modifications.

\section{Conclusions}
In our previous work \cite{Manolakos_paper1, Manolakos_paper2, Manolakos:2022universe, Manolakos:2023hif} the covariant, noncommutative, four-dimensional $dS$ spacetime, fuzzy-$dS_4$, was constructed. In that construction instead of using the algebra of the generators of the relevant isometry group, $SO(1,4)$, we considered a larger one in which the latter could be embedded, following the suggestion of Yang \cite{yang1947} to improve Snyder's model of noncommutativity of coordinates \cite{Snyder:1946qz}. Then using the gauge-theoretic formulation of gravity we have built a gravity model on the fuzzy-$AdS_4$, by gauging the $SO(6) \times U(1) \sim SO(2,4) \times U(1)$.

In the above work, following the route of Snyder's model of noncommutativity of coordinates, and extending the initial group from the isometry group $SO(1,4)$ to $SO(1,5)$, on top of the noncommutativity of coordinates, noncommutativity of momenta and a Heisenberg-type relation were obtained among others. Given the above, we have speculated that a further extension of the isometry group could be even more fruitful. Indeed, our speculation was vindicated, and by extending the isometry group we managed to obtain information not only about the noncommutative space, but also about the group that should be gauged in order to construct a FG theory. It should be noted though, that the presence of a $U(1)$, required for the closing of the anticommutators, remains inevitable as well in the present construction.

In addition, in the present work we have considered the inclusion of fermions in the FG. Although there exist obvious difficulties in this task, namely in introducing chiral fermions in a matrix representation, we managed to overcome them in a straightforward way. As a result we are led to a FG theory based on a noncommutative version of the conformal gravity coupled to a GUT based on $SO(10)$. Attempts to construct FG with $SO(1,5) \times U(1)$ as the gauge group could not be realized, since either the fermions could not become chiral, or the GUT was the minimal $SU(5)$, which has been ruled out phenomenologically. Our plan is to examine further the unification schemes discussed in the present work as well as the one in ref \cite{Roumelioti:2024lvn}, which is the continuum version of the FG examined here, from phenomenological as well as cosmological points of view.

\section*{Acknowledgments}

It is a pleasure to thank Thanassis Chatzistavrakidis, Alex Kehagias and
Pantelis Manousselis for several discussions and suggestions on the content
of the paper. We would like also to thank very much Dieter Lüst for
reading the manuscript and his encouraging comments. D.R. would like to thank NTUA for a fellowship for doctoral studies.
G.Z. would like to thank MPP-Munich, ITP-Heidelberg, and DFG Exzellenzcluster
2181:STRUCTURES of Heidelberg University for their hospitality and support.

\printbibliography

\end{document}